\documentclass[conference]{IEEEtran}
\IEEEoverridecommandlockouts

\usepackage{cite}
\usepackage{amsmath,amssymb,amsfonts}
\usepackage{algorithmic}
\usepackage{graphicx}
\usepackage{textcomp}
\usepackage{xcolor}
\usepackage{amsmath}
\usepackage{listings}
\usepackage{url}
\usepackage{booktabs}
\usepackage{multirow}

\def\BibTeX{{\rm B\kern-.05em{\sc i\kern-.025em b}\kern-.08em
    T\kern-.1667em\lower.7ex\hbox{E}\kern-.125emX}}
\begin{document}

\title{LiteVPNet: A Lightweight Network for Video Encoding Control in Quality-Critical Applications
\thanks{This work was supported by the Horizon CL4 2022, EU Project Emerald, 101119800; ADAPT-SFI Research Centre, Ireland, with Grant ID 13/RC/2106\_P2; and YouTube \& Google Faculty Awards.}
}
\author{\IEEEauthorblockN{Vibhoothi Vibhoothi, Fran\c{c}ois Piti\'{e}, Anil Kokaram}
\IEEEauthorblockA{Sigmedia Group, Department of Electronic and Electrical Engineering, \textit{Trinity College Dublin}, \\ Dublin, Ireland \\
\{vibhootv, pitief, anil.kokaram\}@tcd.ie
}
}

\maketitle

\begin{abstract}
In the last decade, video workflows in the cinema production ecosystem have presented new use cases for video streaming technology. These new workflows, e.g. in On-set Virtual Production, present the challenge of requiring precise quality control and energy efficiency. 
Existing approaches to transcoding often fall short of these requirements, either due to a lack of quality control or computational overhead. 
To fill this gap, we present a lightweight neural network (LiteVPNet) for accurately predicting Quantisation Parameters for NVENC AV1 encoders that achieve a specified VMAF score. 
We use low-complexity features including bitstream characteristics, video complexity measures, and CLIP-based semantic embeddings. 
Our results demonstrate that LiteVPNet achieves mean VMAF errors below 1.2 points across a wide range of quality targets. 
Notably, LiteVPNet achieves VMAF errors within 2 points for over 87\% of our test corpus, c.f. $\approx$61\% with state-of-the-art methods.
LiteVPNet's performance across various quality regions highlights its applicability for enhancing high-value content transport and streaming for more energy-efficient, high-quality media experiences.

\end{abstract}


\section{Introduction}
The streaming paradigm over the internet is well understood and has been driving video coding development for internet use cases for many years. Other applications, especially in cinema production, increasingly require transporting extremely high data volumes with tighter quality constraints than video streaming.
In On-set Virtual Production (OSVP~\cite{hendricks2023rapid}), scenes are filmed on-set with massive LED walls as a backdrop, allowing high-resolution rendering of realistic scenery.
It has rapidly transformed content creation because convincing lighting is implicitly provided by those walls, and the scenery can be adaptively warped to maintain the realism of a camera moving in the corresponding virtual world. The size of the images rendered at high frame rates and high bit depth causes most operations to default to near-lossless encoding at a very high bandwidth that is beyond conventional streaming.
This in turn places a heavy demand on video transport and computational infrastructure, hence increasing energy and resource footprint.

Our recent study~\cite{vibhoothi2025mijvp} demonstrated that modern video codecs (AV1, HEVC with NVENC encoders) achieve perceptually equivalent quality at 2$\times$ to 200$\times$ bitrate savings compared to OSVP industry-standard intermediate codecs, especially after accounting for camera/screen losses due to recording the rendered images in-camera. This presents an opportunity to significantly reduce the data transport and storage requirements between the client and studios.
However, choosing the right encoded representation that guarantees a certain level of quality still requires computing the entire Rate-Distortion curve. That is compute and time-intensive, increasing the latency in just getting the scene set up for filming. The challenge is to generate that representation without many repeated invocations of the encoder in this OSVP workflow, where quality constraints take precedence over bitrate.
While the methods to be discussed were motivated by the demands of OSVP, they apply to any quality-critical workflow, such as remote post-production or high-value content archival.

This paper introduces LiteVPNet, a lightweight network designed to predict Quantisation Parameters for NVENC encoders using low-complexity video features: bitstream characteristics, Video Complexity Analyzer (VCA~\cite{menon2022vca}), and CLIP-based~\cite{radford2021_clip_paper} semantic embeddings.
The model achieves mean VMAF errors ($\Delta$VMAF) of 1 for the test dataset, and over 87\% of the test videos achieved (coverage) $\Delta$VMAF$\le$2, compared to 61\% for existing methods~\cite{mico2023pertilecnndnn,mico2025pertilecnndnn_scenelevel, menon2024jtps}.
This work is one of the first studies demonstrating the viability of advanced codecs for energy-efficient, quality-critical immersive media experiences (VP) in cinema applications. 


\section{Background}
\label{background:rdcurve-est}

A content-adaptive encoder aims to optimise video coding by adjusting the key encoding parameters (e.g., $QP$, $CRF$) based on video characteristics, often via a brute-force analysis or computationally intensive multi-pass encoding.
In 2016, Covell et al.~\cite{2016covellcrfmlyoutube} used neural networks with segment-dependent features to predict CRF values for YouTube's user-generated content (UGC) videos, achieving 80\% of predictions to be under 20\% bitrate error ($\Delta B$). Later~\cite{2018_ieee_rdpred} refined this using a regression model to achieve 90\% coverage within 20\% $\Delta B$ on their test dataset.

More recently, Cai et al.~\cite{2022_qualitypredperclip_bilibili} in 2022 achieved 98.88\% accuracy for CRF parameter prediction to target a  VMAF 91 score.
They used a multi-stage neural network with spatio-temporal features extracted using a fast H.264 NVENC encode combined with clever frame duplication.  They incur 1.55$\times$ overall complexity overhead c.f. a single encode. 
This work used a 500k private video dataset from Bilibili.
In 2023, Mic{\'o}-Engu{\'\i}danos et al.~\cite{mico2023pertilecnndnn, mico2025pertilecnndnn_scenelevel} demonstrated a DNN-based classification network for per-title and per-segment CRF estimation (VP9 with 19 target CRF values) to achieve a target VMAF quality. 
Their feature set (38) comprised statistics extracted using \texttt{ffprobe} and the SI-TI tool at a lower resolution. 
On their test dataset,  a mean deviation of 1.84 VMAF points was reported, with deviations of 2.9, 2.28, 1.35, and 0.83 VMAF points for target VMAFs of 80, 85, 90, and 95, respectively. 
However, the input feature required is computationally intensive and might not be applicable to the OSVP application. We investigate this in Section~\ref{sec:result:compare-methods}.
Furthermore, Menon et al.~\cite{menon2024jtps, 2023_icme_jndmenon2023jasla} have demonstrated that Video Complexity Analyzer (VCA~\cite{menon2022vca}) features along with resolution and target bitrate, can effectively predict CRF for x265 using Random Forest, achieving a high correlation (0.97 $R^2$) and low CRF prediction error (1.87 MAE) to hit desired bitrate targets. However, this method was tested to find CRF for a target bitrate and not target quality.

It is worth observing that Yin et al.~\cite{yin2024contentadaptive_vmaf} have shown that an additional encoding using the predicted $qp$ can drastically improve the quality control~\cite{yin2024contentadaptive_vmaf} at the cost of additional complexity. Other work uses content classification to determine encoding regimes for generating encodes e.g. \cite{ling2020_lecallet_rdcat} or identify an optimal resolution \cite{bhat2020resolutionpred} for achieving a target bitrate. These are not suitable for our use case, which requires finer control.

While eliminating brute-force optimisation, the previous methods primarily target bitrate within a 10-20\% error margin. Approaches targeting quality~\cite{menon2024jtps, mico2023pertilecnndnn} either have computational costs or error margins unsuitable for low-latency OSVP workflows, which need precise quality control.


\section{Low-complexity feature descriptors for videos}
LiteVPNet uses four normalised feature sets extracted from 480×270p downsampled (Lanczos-5) content: A) Bitstream metadata statistics, B) Spatial and Temporal video complexity from VCA, C) Bitstream Information, and D) Semantic Content Features via ``Clippie'' (CPU-based CLIP model). 

{ \textbf{A. Bitstream Features.}}
For a given video, we use NVIDIA Encoder (NVENC) at QP160 using a 40 series Nvidia GPU at Preset-7 without split-frame encoding to generate bitstream-level features. These are extracted with the \texttt{inspect} tool of libaom-av1 encoder.
We employ distributions of bitstream properties at frame-level ($\hat{F}$) and video-level ($\hat{V}$) as features. They include
block sizes ($4\times 4$ up to $128 \times 128$),
transform types/sizes,
skip blocks;
intra block-copy,
palette mode usage,
reference frame types,
use of coding tools (in loop filter etc); 
and bit allocation patterns based on motion vectors.
For copious details, see the project page\footnote{https://github.com/sigmedia/litevpnet \label{projpage}}.


\textbf{B. Bitstream Information.}
We extract bitstream metadata ($\hat{M}$) at the video-level: 
Video duration (secs), 
Bit depth (8-bit or 10-bit),
Average coded quantiser index, 
Video dimension (width and height), and FPS.

\textbf{ C. Video Complexity.} 
To measure spatial and temporal complexity with low computational overhead, we use VCA~\cite{menon2022vca, menon2023vcaagain} ($\hat{A}$). 
They compute texture (spatial complexity, SC) using a lightweight DCT-based energy function and per-pixel SAD values for temporal complexity (TC), alongside frame brightness. Statistical measures (mean, std, min, max, and \{25, 50, 75\%\} percentiles) are extracted separately for I-frames and non-I-frames.

\textbf{D. Clippie.} 
CLIP~\cite{radford2021_clip_paper} is a vision-language model trained on image-text pairs for semantic reasoning with zero-shot transfer capabilities. Clippie~\footnote{clippie, \url{https://github.com/mossblaser/clippie.git}, Access: June 2025} is a CPU-based Numpy implementation of CLIP, producing a 512$\times$1 feature vector for an image input ($\hat{C}$). 
CLIP has proven effective for image recognition~\cite{li2023_clip_imagereg} and is useful for perceptual-quality estimation or as a No-reference quality metric~\cite{2025_ieee_clipvqa, 2024_zeniqa_clip_withprompt}.
Studies indicate that while CLIP features alone provide valuable quality insights, their performance can be substantially enhanced when combined with other features.

All feature sets were normalised to ensure dimensional homogeneity~\cite{2020_datanormalisation}: bitstream metadata to a 0-1 range, VCA features via Min-Max scaling, while CLIP embeddings were already standardised (zero mean and unit variance). The final 754$\times$1 input vector combines bitstream features from the initial 8 frames, full-video VCA metrics, and processed CLIP embeddings.


\section{Dataset}
The dataset comprises 2944 single-shot videos (avg. 300 frames, up to 7s) at various 1080p aspect ratios. To ensure diverse content complexity, videos were sourced from 12 different public and academic collections, including YouTube UGC, Netflix Open-content, AOM-CTC, Xiph.org, the SJTU Dataset, ASC SteM2, and the Inter4K dataset, among others. The dataset was split 80-20\% for training and testing, ensuring no content overlap. Please check the project page\footref{projpage} for more details.

\section{LiteVPNet Network Architecture}
\begin{figure}
    \centering
    \begin{tabular}{c}
    \includegraphics[width=0.9\columnwidth]{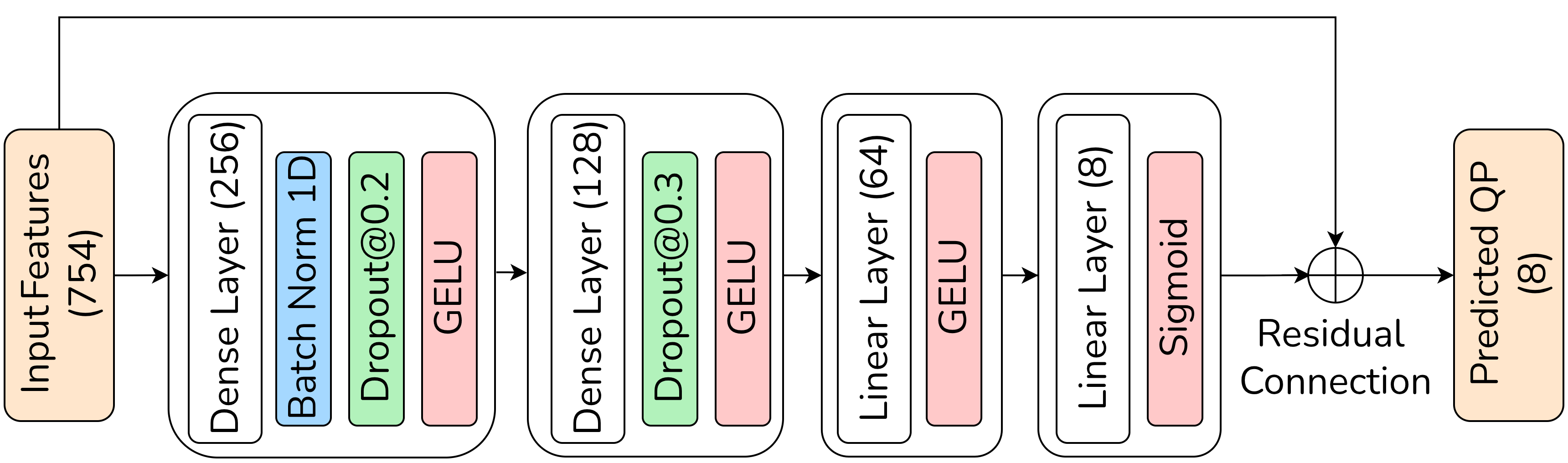} \\
    (a) \\
    \includegraphics[width=0.9\columnwidth]{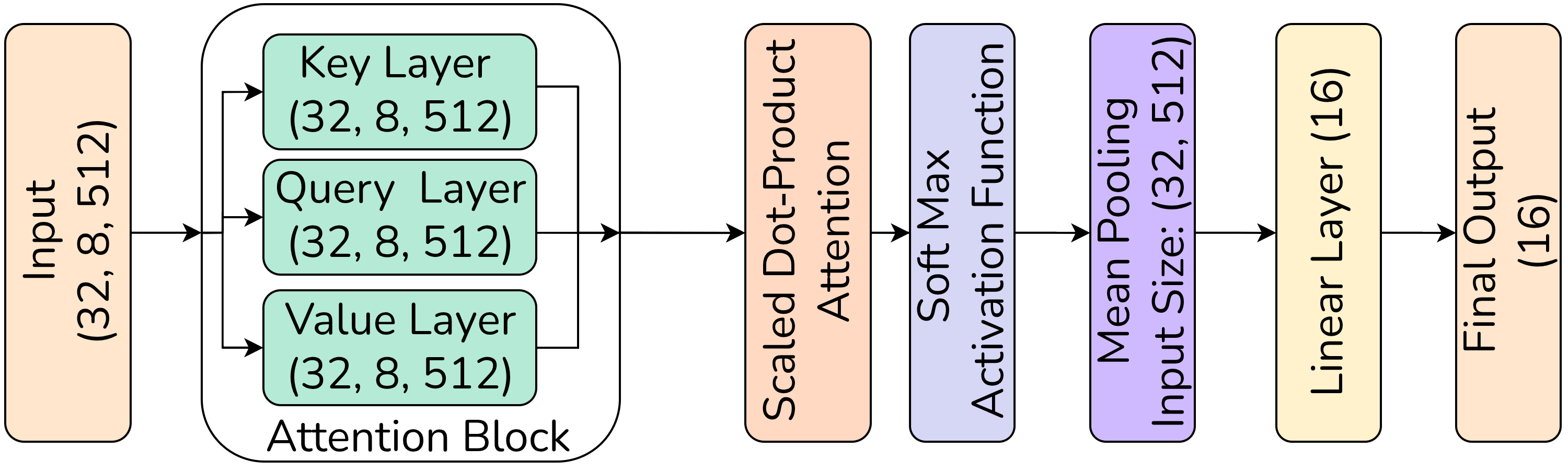} \\
    (b)
    \end{tabular}          
    \caption{Network Design for (a) LiteVPNet Network, (b) ClipNet feature embedding network.}
    \label{fig:dnn-qppreddnn}
    \vspace{-0.7em}
\end{figure}

LiteVPNet employs a lightweight DNN architecture (Figure~\ref{fig:dnn-qppreddnn}) to predict optimal QP for a given video to meet a desired quality target. 
To address the different quality requirements for different scenarios, from visually lossless VMAF 99 for virtual production backdrops to VMAF 91 for high-quality streaming. 
We define the ground truth $QP$ and VMAF pairs through an exhaustive 24 $QP$ encodes for each video. 
PCHIP interpolation is then used to get eight distinct target QPs for the target VMAF scores of \{99, 97, 95, 91, 88, 85, 83, 80\}.

The architecture consists of two jointly trained components, as depicted in Figure~\ref{fig:dnn-qppreddnn}. 
First, ClipNet, a Transformer-style attention network processes the 4096-dimensional Clippie feature vector (derived from 8 frames, 512$\times$8=4096 vector). This network uses self-attention and a linear projection layer to produce a compact 16$\times$1 embedding. 
This embedding is then concatenated with the VCA and bitstream features to form the 754-dimensional input for the main LiteVPNet DNN. 
The LiteVPNet is a feed-forward network with four fully-connected layers (754\(\rightarrow\)256\(\rightarrow\)128\(\rightarrow\)64\(\rightarrow\)8), using Batch Normalisation, GELU activation, dropouts, and residual connections to improve gradient flow with sigmoid activation function. The output is a normalised QP value for the eight VMAF targets.

The model was implemented in PyTorch and trained using a custom \texttt{TolerantWeightedMSELoss} loss function, which applies no penalty for predictions within \(\pm2\) VMAF points as tolerance with a batch size of 32. 
The final loss \(\mathcal{L}\) combines the L1 loss on the predicted QP values for $t$ targets ($\mathcal{L}_{a} = || \text{LiteVPNet}(X) - QP_t||_1$) and another L1 loss on the resulting VMAF scores ($\mathcal{L}_{b}= \alpha  || \text{VMAF}(QP_p) - \text{VMAF}(QP_t)||_1$), where the  \(\alpha\) is set to be 1.  The final loss function is $\mathcal{L} = \mathcal{L}_{a} + \mathcal{L}_{b}$. 
During training, the VMAF scores for the dataset were pre-computed (interpolated with PCHIP to cover the full range). 
We used the Adam optimiser (Learning Rate=\(1e^{-4}\)) with L2 regularisation (\(1e^{-5}\)) and a \texttt{ReduceLROnPlateau} scheduler. 
The complete model, with 242k parameters for the LiteVPNet and 798k for the ClipNet network, converged after 103 epochs on an NVENC 40 series GPU. 
We observed that jointly training the ClipNet yielded better $R^2$ and performance as opposed to training individually.



\section{Results}

\begin{figure*}
    \centering
    \begin{tabular}{cc}
    \includegraphics[width=0.7\columnwidth]{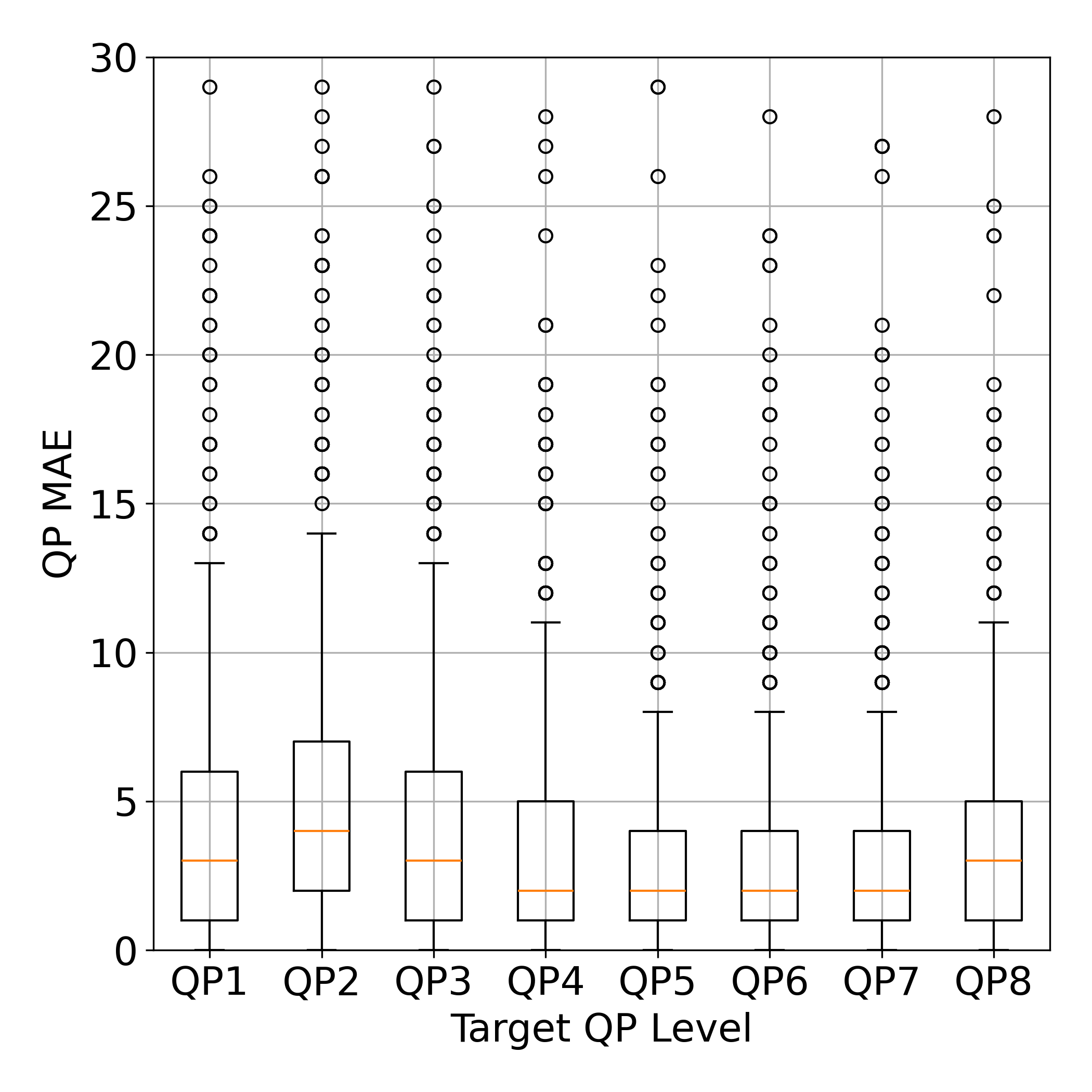} &
    \includegraphics[width=0.7\columnwidth]{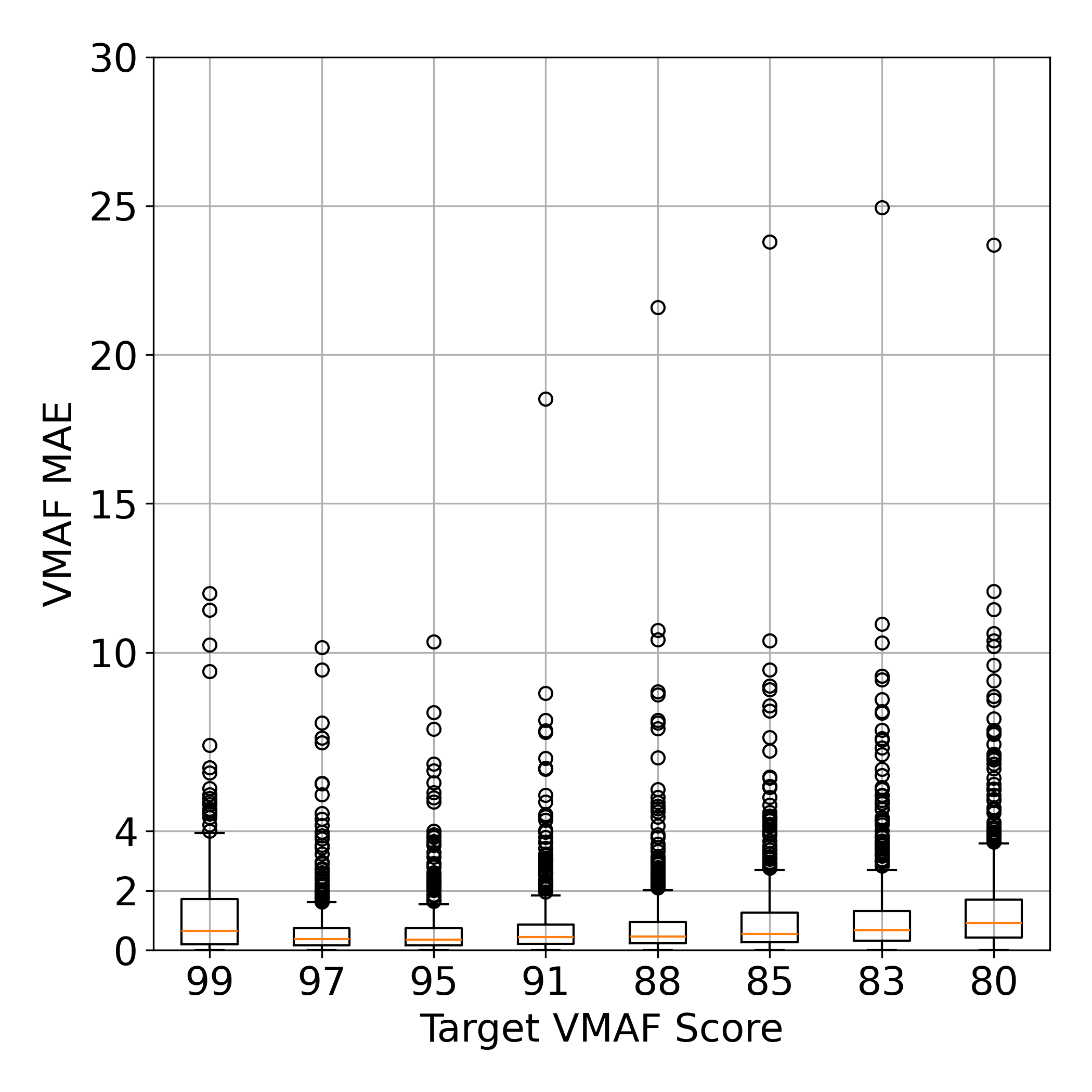} \\
    (a) & (b) 
    \end{tabular}
    \caption{Boxplots (25\% and 75\% percentile)  of QP and VMAF prediction errors (y-axis). Each target $QP_i$ (x-axis) corresponds to a VMAF level from 99 down to 80. The average VMAF MAE is 1.0, demonstrating good quality control despite an average QP MAE of 4.5.}

    \label{fig:qppred-qp-mae}
    \vspace{-0.8em}

\end{figure*}

\begin{figure}
    \centering
    \includegraphics[width=0.9\columnwidth]{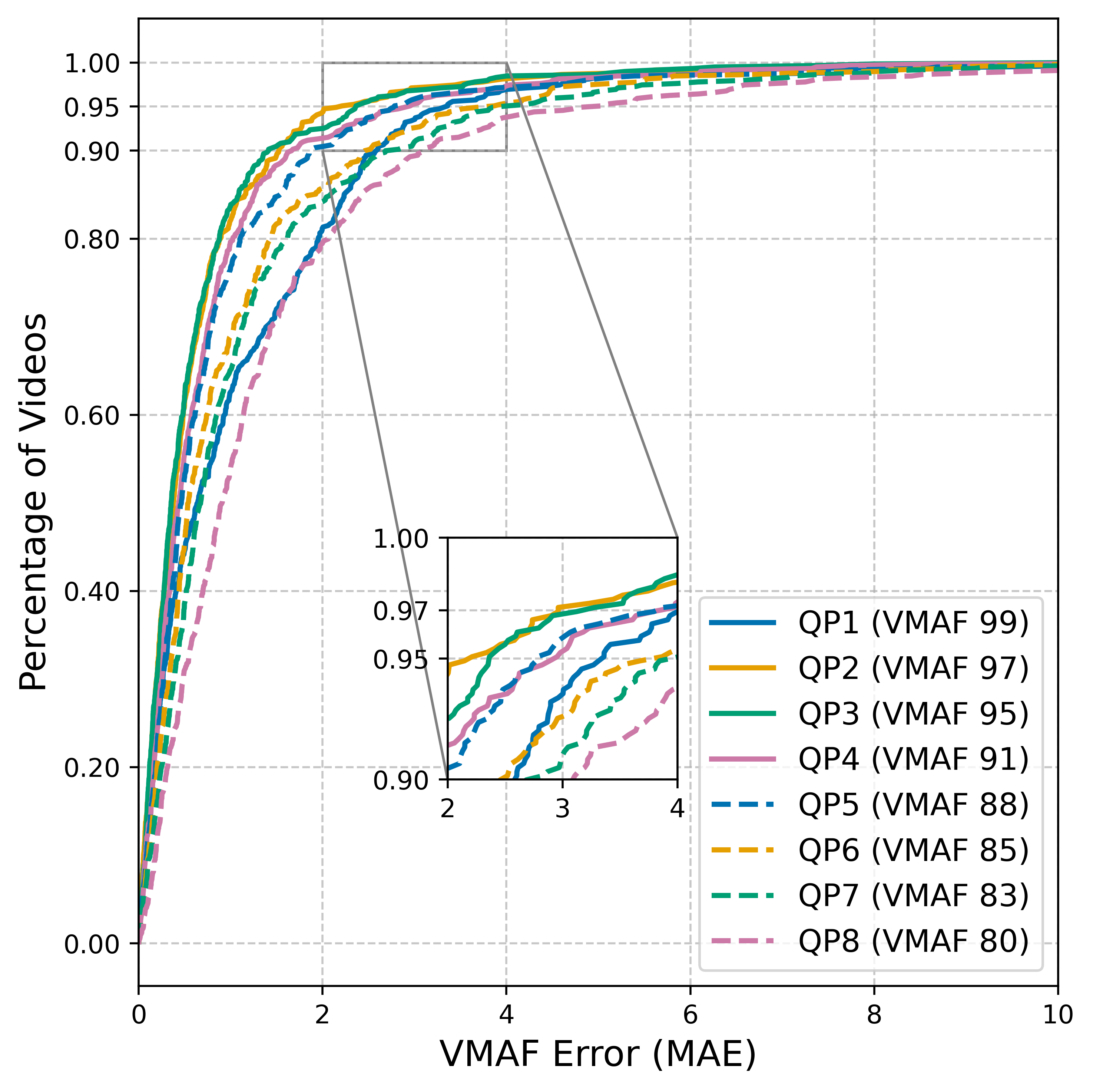}
    \caption{
    CDF of VMAF errors for different target VMAF levels (99\(\cdots\)80). The x-axis indicates the absolute VMAF error, and the y-axis represents the percentage of videos within that error threshold. For example, the QP2 target (VMAF 97) achieves 97\% dataset coverage with a VMAF error of \(\le3\).
    }
    \vspace{-1em}
    \label{fig:qppred-vmaf-cdfdis}
\end{figure}

Figure~\ref{fig:qppred-qp-mae} shows the Mean Absolute Error (MAE) of QP and VMAF predictions for different quality targets. 
The average QP prediction error is 4.5 (median 2.5, standard deviation ($\sigma$) 6.5).
Given the AV1 QP range of 0-255 ($\approx$4 QP values per QStep), these error margins are acceptable. 
The error in VMAF is smaller than QP, with mean errors consistently below 1.5 (average 1.0, median 0.5,  $\sigma$ 1.5). 
Notably, only 4.9\% of videos had  $\Delta$VMAF$\ge 6$ (1 JND), and only 1 video with $\Delta$VMAF$\ge12$. These videos had complex elements like high noise/grain, banding, and smoke.

Figure~\ref{fig:qppred-vmaf-cdfdis} presents the VMAF error CDF for all target quality levels, showing the percentage of videos achieving specific error thresholds. 
All quality levels (QP1-QP8) demonstrate similar behaviour. Key observations from the CDF distribution include: 
i) All quality levels (QP1-QP8) achieve approximately 80\% dataset coverage for VMAF errors \(\le2\). 
ii) With an acceptable VMAF error range of \(\le4\) points, over 93\% of the dataset can be accurately predicted using the model.

We observe a non-linear relationship between QP and VMAF error. Large errors in QP (median 2-4) do not imply large errors in VMAF (including other metrics like PSNR and SSIM). This characteristic is fundamental to rate-distortion optimisation, and a fortunate outcome because it is the VMAF error which is most important in this use case.

\subsection{Ablation Study of the LiteVPNet Input Features}

An ablation study was conducted to evaluate the individual contributions of LiteVPNet's distinct feature categories. This involved systematically removing: FrameLevelStats (\(\hat{F}\)), VideoLevelStats (\(\hat{V}\)), VideoBitstreamMetadata (\(\hat{M}\)), VCA (\(\hat{A}\)), and ClippieEmbeddings (\(\hat{C}\)). 
The performance of the full LiteVPNet model, incorporating all feature sets, served as the baseline for comparison.

Table~\ref{tab:ablationstudy} presents the QP and VMAF MAE results, alongside the percentage coverage for VMAF errors \(\le2\) and \(\le4\), across three quality bands: High (\(QP1{\cdots}QP4\), VMAF \(91\cdots99\)), Medium (\(QP3{\cdots}QP6\), VMAF \(85\cdots95\)), and Low (\(QP5{\cdots}QP8\), VMAF \(80\cdots88\)). The full LiteVPNet model (baseline) achieves the best overall performance, with mean MAE values of 4.5 for QP and 1.0 for VMAF, and coverage of 87.3\% and 96.5\% for \(VMAF_{MAE} \le 2\) and \(VMAF_{MAE} \le 4\), respectively.
Removing ClippieEmbeddings (\(\hat{C}\)) has the most significant performance degradation: QP MAE increases to 7.2 (from 4.5), VMAF MAE rises to 1.5 (from 1.0), and coverage for \(VMAF_{MAE} \le 2\) drops to 75.6\% (from 87.3\%). 
VCA removal (\(\hat{A}\)) also substantially impacts performance, increasing QP MAE to 6.1 and VMAF MAE to 1.2. 
Other feature removals exhibit more modest effects; for instance, excluding VideoLevelStats (\(\hat{V}\)) results in a QP MAE of 4.9 while the VMAF MAE remains unchanged at 1.0. VideoBitstreamMetadata removal (\(\hat{M}\)) shows similar minor impacts. This confirms that ClippieEmbeddings and VCA are the most influential components of LiteVPNet's predictive capability.

\begin{table*}
\centering
\caption{
Ablation study of LiteVPNet. The table shows QP/VMAF MAE and VMAF error coverage (\(\le2\) and \(\le4\)) when removing feature sets across three quality ranges: Low (VMAF 80 to 88), Medium (VMAF 85 to 95), and High (VMAF 91 to 99). The full model (bold) is the best performer. Removing Clippie embeddings (\(\hat{C}\)) causes the most significant performance degradation. (Features: \(\hat{F}\) = FrameLevelStats, \(\hat{V}\) = VideoLevelStats, \(\hat{M}\) = VideoBitstreamMetadata,  \(\hat{A}\) = VCA, and \(\hat{C}\) = ClippieEmbeddings).
}
\label{tab:ablationstudy}
\begin{tabular}{@{}lllllllllllllllllll@{}}
\toprule
\multirow{2}{*}{\textbf{Method}} & 
\multicolumn{4}{c}{\textbf{MAE of QP ($\downarrow$)}} & \multicolumn{4}{c}{\textbf{MAE of VMAF ($\downarrow$)}} & \multicolumn{4}{c}{\textbf{$\text{$\Delta$VMAF}<=2$\% ($\uparrow$)}} & 
\multicolumn{4}{c}{\textbf{$\text{$\Delta$VMAF}<=4$\% ($\uparrow$)}} \\ 
\cmidrule(l){2-5} \cmidrule(l){6-9} \cmidrule(l){10-13} \cmidrule(l){14-17}
 & \textbf{$\overline{X}$} & \textbf{Low} & \textbf{Med} & \textbf{High}  & \textbf{$\overline{X}$} & \textbf{Low} & \textbf{Med} & \textbf{High}  & \textbf{$\overline{X}$} & \textbf{Low} & \textbf{Med} & \textbf{High} & \textbf{$\overline{X}$} & \textbf{Low} & \textbf{Med} & \textbf{High} \\ \midrule
\textbf{LiteVPNet} & \textbf{4.5} & \textbf{3.9} & \textbf{4.2} & \textbf{5.1} & \textbf{1.0} & \textbf{1.2} & \textbf{0.9} & \textbf{0.8} & \textbf{87.3} & \textbf{84.9} & \textbf{89.9} & \textbf{89.8} & \textbf{96.5} & \textbf{95.3} & 97.0 & \textbf{97.7} \\
LiteVPNet - $\hat{F}$ & 5.7 & 5.0 & 5.4 & 6.4 & 1.2 & 1.4 & 1.1 & 0.9 & 83.5 & 78.3 & 86.4 & 88.7 & 95.2 & 93.2 & 95.9 & 97.2 \\
LiteVPNet - $\hat{V}$ & 4.9 & 4.2 & 4.6 & 5.5 & 1.0 & 1.3 & 0.9 & 0.8 & 86.3 & 82.3 & \textbf{89.7} & \textbf{90.2} & 96.3 & \textbf{94.6} & \textbf{97.2} & \textbf{98.1} \\
LiteVPNet - $\hat{M}$ & 5.0 & 4.2 & 4.6 & 5.8 & 1.1 & 1.3 & 0.9 & 0.9 & 85.9 & 82.6 & 89.0 & 89.3 & 95.9 & 94.3 & 96.7 & 97.5 \\
LiteVPNet - $\hat{A}$ & 6.1 & 5.1 & 5.6 & 7.0 & 1.2 & 1.5 & 1.1 & 1.0 & 83.4 & 79.0 & 86.9 & 87.8 & 93.9 & 91.3 & 94.5 & 96.4 \\
LiteVPNet - $\hat{C}$ & 7.2 & 6.8 & 7.2 & 7.7 & 1.5 & 2.0 & 1.4 & 1.1 & 75.6 & 65.4 & 78.3 & 85.7 & 92.9 & 88.6 & 94.2 & 97.2 
\\ \bottomrule

\end{tabular}%
\end{table*}

\subsection{Comparison with other methods}
\label{sec:result:compare-methods}
This section compares LiteVPNet against Mico-DNN~\cite{mico2023pertilecnndnn} and JTPS~\cite{menon2024jtps}, evaluating all methods on the same dataset with consistent train-test splits ensured by identical random seeds. 
The target is AV1 QP value prediction using the NVENC encoder, enabling direct performance comparison.

Mico-DNN extracts features via \texttt{ffprobe}, including features such as hue, saturation, luminance, chrominance, normalised grey-level entropy per channel (Y, UV), and spatial and temporal information. 
For this comparison, Mico-DNN's feature extraction process was applied to our dataset, including video normalisation to x264 CRF 0  with ultrafast preset and downsampling to 240p with Lanczos filtering. 
JTPS uses VCA metrics and video metadata (resolution, frame rate) to predict CRF values for a given target bitrate. 
One key change which was adopted to the JTPS method was adjusting the model to input $QP$ as opposed to bitrate for predicting target VMAF, aligning with LiteVPNet's objective. All three JTPS regression models (Linear Regression, XGBoost, and Random Forest) hyperparameters were re-tuned.

Table~\ref{tab:compare-table-methods} shows the MAE for QP and VMAF, presented similarly to Table~\ref{tab:ablationstudy}. 
LiteVPNet offers far superior quality control. 
It consistently achieves the lowest MAE across all categories: Mean QP MAE is 4.5, significantly outperforming JTPS (13) and Mico-DNN (33.7). 
Similarly, VMAF MAE of LiteVPNet (1.0) surpasses JTPS (2.1) and Mico-DNN (5.9). 
For video coverage for \(VMAF_{MAE} \le 2\), LiteVPNet covers 87.3\% of videos, better than JTPS (61.1\%) and Mico-DNN (27.8\%). For \(VMAF_{MAE} \le 4\), LiteVPNet achieves 96.5\% coverage, outperforming JPTS (87.1\%) and Mico-DNN (49.6\%). 
These results confirm greater reliability and precision in perceptual quality control of LiteVPNet.

\begin{table*}
\centering
\caption{Comparison of Different QP Prediction Methods. Mean Absolute Error (MAE) for QP and VMAF prediction, alongside percentage coverage for VMAF errors \(\le2\) and \(\le4\). Results are presented as overall mean (\(\overline{X}\)) and across three quality ranges: Low (VMAF 80 to 88), Medium (VMAF 85 to 95), and High (VMAF 91 to 99). The proposed LiteVPNet consistently outperforms Mico-DNN and JTPS.}
\label{tab:compare-table-methods}
\begin{tabular}{@{}lrrrrrrrrrrrrrrrrrr@{}}
\toprule
\multirow{2}{*}{\textbf{Method}} & 
\multicolumn{4}{l}{\textbf{MAE of QP ($\downarrow$)}} & \multicolumn{4}{l}{\textbf{MAE of VMAF ($\downarrow$)}} & \multicolumn{4}{l}{\textbf{$\Delta\text{VMAF}<=2\%$ ($\uparrow$)}} & 
\multicolumn{4}{l}{\textbf{$\text{$\Delta$VMAF}<=4\%$ ($\uparrow$)}} \\ 
\cmidrule(l){2-5} \cmidrule(l){6-10} \cmidrule(l){10-13} \cmidrule(l){14-17}  
 & \multicolumn{1}{l}{\textbf{$\overline{X}$}} & \multicolumn{1}{l}{\textbf{Low}} & \multicolumn{1}{l}{\textbf{Med}} & \multicolumn{1}{l}{\textbf{High}}  & 
 \multicolumn{1}{l}{\textbf{\textbf{$\overline{X}$}}} & \multicolumn{1}{l}{\textbf{Low}} & \multicolumn{1}{l}{\textbf{Med}} & \multicolumn{1}{l}{\textbf{High}} &  
 \multicolumn{1}{l}{\textbf{$\overline{X}$}} & \multicolumn{1}{l}{\textbf{Low}} & \multicolumn{1}{l}{\textbf{Med}} & \multicolumn{1}{l}{\textbf{High}} & \multicolumn{1}{l}{\textbf{\textbf{$\overline{X}$}}} & \multicolumn{1}{l}{\textbf{Low}} & \multicolumn{1}{l}{\textbf{Med}} & \multicolumn{1}{l}{\textbf{High}} \\ \midrule
Mico-DNN~\cite{mico2023pertilecnndnn} & 33.7 & 26.3 & 32.3 & 41.1 & 5.9 & 7.6 & 6.3 & 4.3 & 27.8 & 16.9 & 22.3 & 38.6 & 49.6 & 34.9 & 43.4 & 64.4 \\
JTPS~\cite{menon2024jtps} & 13 & 9.5 & 11.1 & 16.6 & 2.1 & 2.5 & 1.6 & 1.6 & 61.1 & 53.6 & 62.1 & 68.5 & 87.1 & 80.4 & 88.4 & 93.8 \\
LiteVPNet (Ours) & \textbf{4.5} & \textbf{3.9} & \textbf{4.2} & \textbf{5.1} & \textbf{1.0} & \textbf{1.2} & \textbf{0.9} & \textbf{0.8} & \textbf{87.3} & \textbf{84.9} & \textbf{89.9} & \textbf{89.8} & \textbf{96.5} & \textbf{95.3} & \textbf{97.0} & \textbf{97.7} \\ \bottomrule
\end{tabular}%
\end{table*}

\subsection{Computational Complexity}
End-to-end runtime was benchmarked on all shots from two 1080p Netflix short films (Meridan: 12mins, NocturneRoom: 11mins, average shot length: 9.5s). This includes downsampling, feature extraction, and prediction. Our method, LiteVPNet, processed each shot in approximately 3.0s, outperforming JTPS (5.6s, 1.9$\times$) and Mico-DNN (5.3s, 1.7$\times$). This efficiency is further highlighted when compared against a traditional brute-force approach used for initial parameter estimation, which requires multiple encodings (8 QPs, 17s) and VMAF computation (180s), resulting in a 65$times$ speed-up. 
Although LiteVPNet has more parameters (242k + 742k, 984k) than Mico-DNN (28k), the model inference time is the same (0.28s). This is due to the parallelisation of the neural networks. JPTS achieves the fastest inference time of 0.11s due to the Random
Forest regression model. This confirms its suitability for latency-sensitive production workflows.

\section{Conclusion}
This work introduced LiteVPNet to predict QP parameters to achieve specific perceptual quality targets in AV1 encoding. 
The system is an efficient neural network that combines diverse feature sets, including bitstream statistics, VCA, and semantic CLIP embeddings, processed via a self-attention-based model. 
It achieves high performance, demonstrated by a low mean QP MAE of 4.5 and a mean VMAF MAE of 1.0. Furthermore, our results in a sense highlight the non-linearity inherent in R/D optimisation: moderate QP variations (errors of 3.9-5.1) yield considerably smaller VMAF errors (0.8-1.2), indicating precise perceptual control. 
Future work will expand support to UHD/HDR content and validate the model on more OSVP-specific datasets to enhance its practical applicability.

\vspace{-0.2em}

\bibliographystyle{IEEEtran}
\bibliography{references}

\begin{thebibliography}{10}
\providecommand{\url}[1]{#1}
\csname url@samestyle\endcsname
\providecommand{\newblock}{\relax}
\providecommand{\bibinfo}[2]{#2}
\providecommand{\BIBentrySTDinterwordspacing}{\spaceskip=0pt\relax}
\providecommand{\BIBentryALTinterwordstretchfactor}{4}
\providecommand{\BIBentryALTinterwordspacing}{\spaceskip=\fontdimen2\font plus
\BIBentryALTinterwordstretchfactor\fontdimen3\font minus \fontdimen4\font\relax}
\providecommand{\BIBforeignlanguage}[2]{{%
\expandafter\ifx\csname l@#1\endcsname\relax
\typeout{** WARNING: IEEEtran.bst: No hyphenation pattern has been}%
\typeout{** loaded for the language `#1'. Using the pattern for}%
\typeout{** the default language instead.}%
\else
\language=\csname l@#1\endcsname
\fi
#2}}
\providecommand{\BIBdecl}{\relax}
\BIBdecl

\bibitem{hendricks2023rapid}
R.~Hendricks, ``{Rapid Industry Solutions (RiS) Initiative, On-Set Virtual Production (OSVP), and Open Services Alliance (OSA) at SMPTE},'' \emph{SMPTE Motion Imaging Journal}, vol. 132, no.~8, pp. 78--80, 2023.

\bibitem{vibhoothi2025mijvp}
V.~Vibhoothi, J.~Zouein, F.~Piti{\'e}, C.~Nash, J.~Bentley, P.~Coulam-Jones, and A.~Kokaram, ``Demystifying the use of compression in virtual production,'' in \emph{Motion Imaging Journal}, vol. 133, 05 2025, pp. 44--54.

\bibitem{menon2022vca}
V.~V. Menon, C.~Feldmann, H.~Amirpour, M.~Ghanbari, and C.~Timmerer, ``Vca: video complexity analyzer,'' in \emph{Proceedings of the 13th ACM multimedia systems conference}, 2022, pp. 259--264.

\bibitem{radford2021_clip_paper}
A.~Radford, J.~W. Kim, C.~Hallacy, A.~Ramesh, G.~Goh, S.~Agarwal, G.~Sastry, A.~Askell, P.~Mishkin, J.~Clark \emph{et~al.}, ``Learning transferable visual models from natural language supervision,'' in \emph{International conference on machine learning}.\hskip 1em plus 0.5em minus 0.4em\relax PmLR, 2021, pp. 8748--8763.

\bibitem{mico2023pertilecnndnn}
F.~Mic{\'o}-Engu{\'\i}danos, W.~Moina-Rivera, J.~Guti{\'e}rrez-Aguado, and M.~Garcia-Pineda, ``Per-title and per-segment crf estimation using dnns for quality-based video coding,'' \emph{Expert Systems with Applications}, vol. 227, p. 120289, 2023.

\bibitem{mico2025pertilecnndnn_scenelevel}
\BIBentryALTinterwordspacing
F.~M. Mico-Enguidanos, J.~Gutierrez-Aguado, and M.~Garcia-Pineda, ``Per title video quality encoding with crf estimation based on scenes using dnn,'' in \emph{Proceedings of the 12th Euro American Conference on Telematics and Information Systems}, ser. EATIS 2024.\hskip 1em plus 0.5em minus 0.4em\relax New York, NY, USA: Association for Computing Machinery, 2025. [Online]. Available: \url{https://doi.org/10.1145/3685243.3685280}
\BIBentrySTDinterwordspacing

\bibitem{menon2024jtps}
V.~V. Menon, P.~T. Rajendran, C.~Feldmann, K.~Schoeffmann, M.~Ghanbari, and C.~Timmerer, ``Jnd-aware two-pass per-title encoding scheme for adaptive live streaming,'' \emph{IEEE Transactions on Circuits and Systems for Video Technology}, vol.~34, no.~2, pp. 1281--1294, 2024.

\bibitem{2016covellcrfmlyoutube}
M.~Covell, M.~Arjovsky, Y.-c. Lin, and A.~Kokaram, ``Optimizing transcoder quality targets using a neural network with an embedded bitrate model,'' \emph{Electronic Imaging}, vol.~28, pp. 1--7, 2016.

\bibitem{2018_ieee_rdpred}
Y.~Sun, M.~Jin, L.~Li, and Z.~Li, ``A machine learning approach to accurate sequence-level rate control scheme for video coding,'' in \emph{2018 25th IEEE International Conference on Image Processing (ICIP)}, 2018, pp. 1013--1017.

\bibitem{2022_qualitypredperclip_bilibili}
C.~Cai, Y.~Wang, X.~Li, and T.~Ye, ``Quality-constant per-shot encoding by two-pass learning-based rate factor prediction,'' in \emph{2022 IEEE International Conference on Visual Communications and Image Processing (VCIP)}, 2022, pp. 1--1.

\bibitem{2023_icme_jndmenon2023jasla}
V.~V. Menon, J.~Zhu, P.~T. Rajendran, H.~Amirpour, P.~Le~Callet, and C.~Timmerer, ``Just noticeable difference-aware per-scene bitrate-laddering for adaptive video streaming,'' in \emph{2023 IEEE International Conference on Multimedia and Expo (ICME)}.\hskip 1em plus 0.5em minus 0.4em\relax IEEE, 2023, pp. 1673--1678.

\bibitem{yin2024contentadaptive_vmaf}
S.~Yin, Z.~Zhang, P.~Ning, Q.~Chen, J.~Chen, Q.~Zhou, G.~Lu, and L.~Song, ``Content-adaptive rate-quality curve prediction model in media processing system,'' in \emph{2024 IEEE International Conference on Visual Communications and Image Processing (VCIP)}.\hskip 1em plus 0.5em minus 0.4em\relax IEEE, 2024, pp. 1--5.

\bibitem{ling2020_lecallet_rdcat}
S.~Ling, Y.~Baveye, P.~Le~Callet, J.~Skinner, and I.~Katsavounidis, ``Towards perceptually-optimized compression of user generated content (ugc): Prediction of ugc rate-distortion category,'' in \emph{2020 IEEE international conference on multimedia and expo (ICME)}.\hskip 1em plus 0.5em minus 0.4em\relax IEEE, 2020, pp. 1--6.

\bibitem{bhat2020resolutionpred}
M.~Bhat, J.-M. Thiesse, and P.~Le~Callet, ``A case study of machine learning classifiers for real-time adaptive resolution prediction in video coding,'' in \emph{2020 IEEE International Conference on Multimedia and Expo (ICME)}.\hskip 1em plus 0.5em minus 0.4em\relax IEEE, 2020, pp. 1--6.

\bibitem{menon2023vcaagain}
V.~V. Menon, C.~Feldmann, K.~Schoeffmann, M.~Ghanbari, and C.~Timmerer, ``Green video complexity analysis for efficient encoding in adaptive video streaming,'' in \emph{Proceedings of the First International Workshop on Green Multimedia Systems}, 2023, pp. 16--18.

\bibitem{li2023_clip_imagereg}
X.~Li, D.~Lian, Z.~Lu, J.~Bai, Z.~Chen, and X.~Wang, ``Graphadapter: Tuning vision-language models with dual knowledge graph,'' \emph{Advances in Neural Information Processing Systems}, vol.~36, pp. 13\,448--13\,466, 2023.

\bibitem{2025_ieee_clipvqa}
F.~Xing, M.~Li, Y.-G. Wang, G.~Zhu, and X.~Cao, ``Clipvqa: Video quality assessment via clip,'' \emph{IEEE Transactions on Broadcasting}, vol.~71, no.~1, pp. 291--306, 2025.

\bibitem{2024_zeniqa_clip_withprompt}
T.~Miyata, ``Zen-iqa: Zero-shot explainable and no-reference image quality assessment with vision language model,'' \emph{IEEE Access}, vol.~12, pp. 70\,973--70\,983, 2024.

\bibitem{2020_datanormalisation}
J.~Shao, K.~Hu, C.~Wang, X.~Xue, and B.~Raj, ``Is normalization indispensable for training deep neural networks?'' in \emph{Proceedings of the 34th International Conference on Neural Information Processing Systems}, ser. NIPS '20.\hskip 1em plus 0.5em minus 0.4em\relax Red Hook, NY, USA: Curran Associates Inc., 2020.

\end{thebibliography}

\end{document}